\documentstyle[sprocl,epsf]{article}

\def\hMpc{\ifmmode{h^{-1}{\rm Mpc}}\else{$h^{-1}$Mpc}\fi}

\begin{document}

\title{ Regularity in the distribution of superclusters }

\author{ Martin Kerscher }

\address{ Sektion Physik, Ludwig--Maximilians Universit\"{a}t, \\
Theresienstr. 37, D-80333 M\"{u}nchen\\
email: kerscher@stat.physik.uni-muenchen.de }

\maketitle

\abstracts{ Using a measure of clustering derived from the nearest
neighbour distribution and the void probability function we are able to
distinguish between regular and clustered structures.  With an example
we show that regularity is a property of a point set, which may be
invisible in the two point correlation function.  Applying this
measure to a supercluster catalogue~{}\cite{einasto:supercluster_data}
we conclude that there is some evidence for regular structures on
large scales.  }

\section{Introduction}

Recently Einasto et al.{}\cite{einasto:120mpc} claimed that the
superclusters form a quasi--regular network.  A similar periodicity in
the galaxy distribution with a period of 128\hMpc\ ($H = 100 h {\rm
km/s / Mpc}$) was reported earlier from an analysis of pencil
beams~{}\cite{broadhurst:large-scale} but the statistical significance
was regarded as weak. Actually this periodicity may be explained
within gravitational clustering from Gaussian initial
conditions~{}\cite{weiss:highres}; see however the discussion by
Szalay~{}\cite{szalay:walls}.

\section{Method}

To analyze the distribution of points given by the redshift
coordinates of the superclusters~{}\cite{einasto:supercluster_data} we
use the spherical contact distribution $F(r)$, i.e.\ the {\em
distribution function of the distance $r$ of an arbitrary point to the
nearest point of the process}.  $F(r)$ is equal to the expected
fraction of volume occupied by points which are not farther than $r$
from the nearest point of the process. Therefore, $1-F(r)$ is equal to
the void probability function $P_0(r)$.
As another tool we use the nearest neighbour distribution $G(r)$ which is
defined as the {\em distribution function of distances $r$ of a point
of the process to the nearest other point of the process}.
For a homogeneous Poisson process the probability to find a point only
depends on the mean number density $\overline{\rho}$, leading to the
well--known result
\begin{equation} \label{eq:F_poi}
 F(r) = 1 - \exp\left(- \overline{\rho} \frac{4 \pi}{3} r^3\right) = G(r).
\end{equation}
\begin{figure}
 \begin{center}
 \epsfxsize=3.9cm
 \begin{minipage}{\epsfxsize} \epsffile{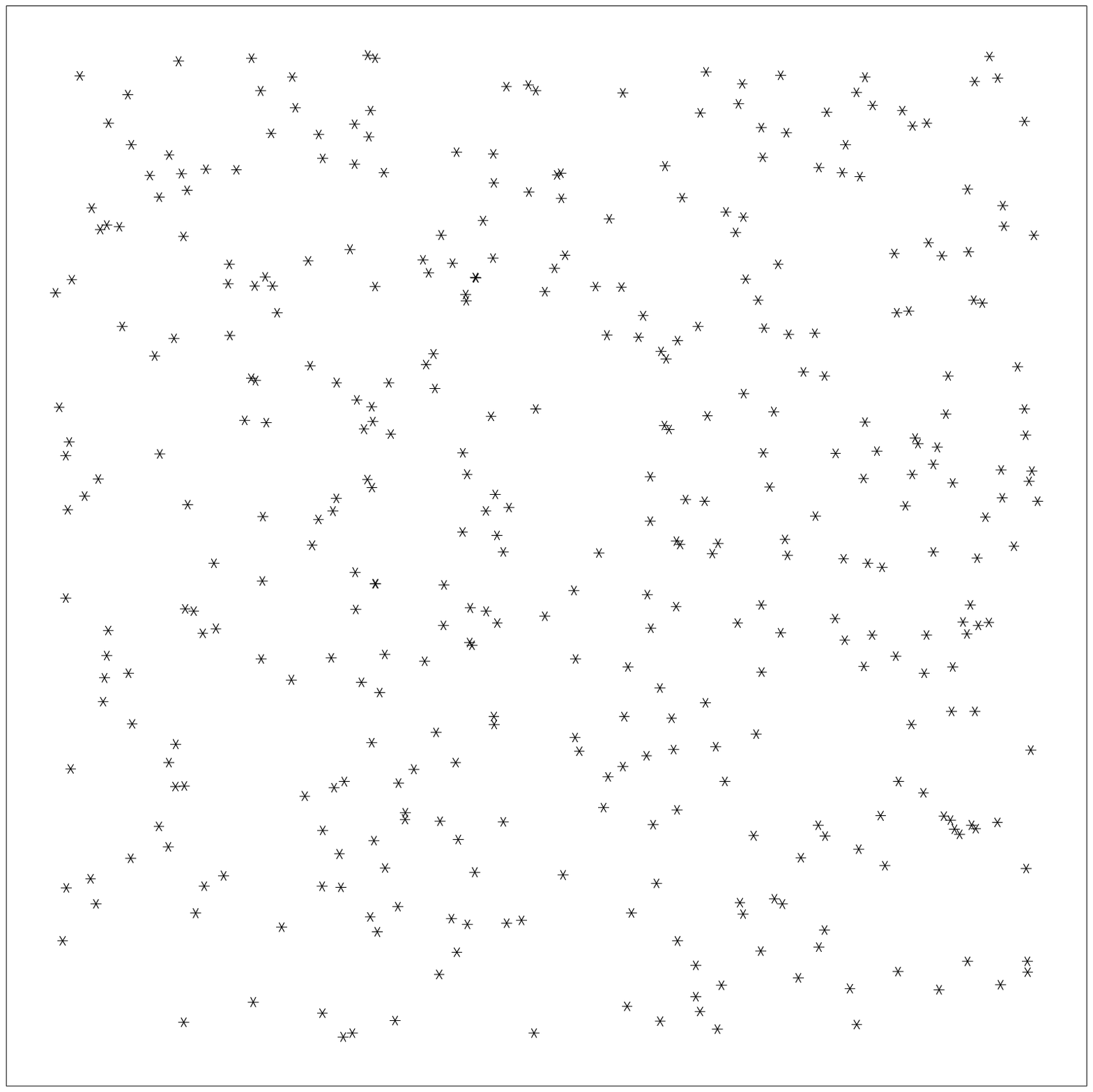} \end{minipage} 
 \epsfxsize=3.9cm
 \begin{minipage}{\epsfxsize} \epsffile{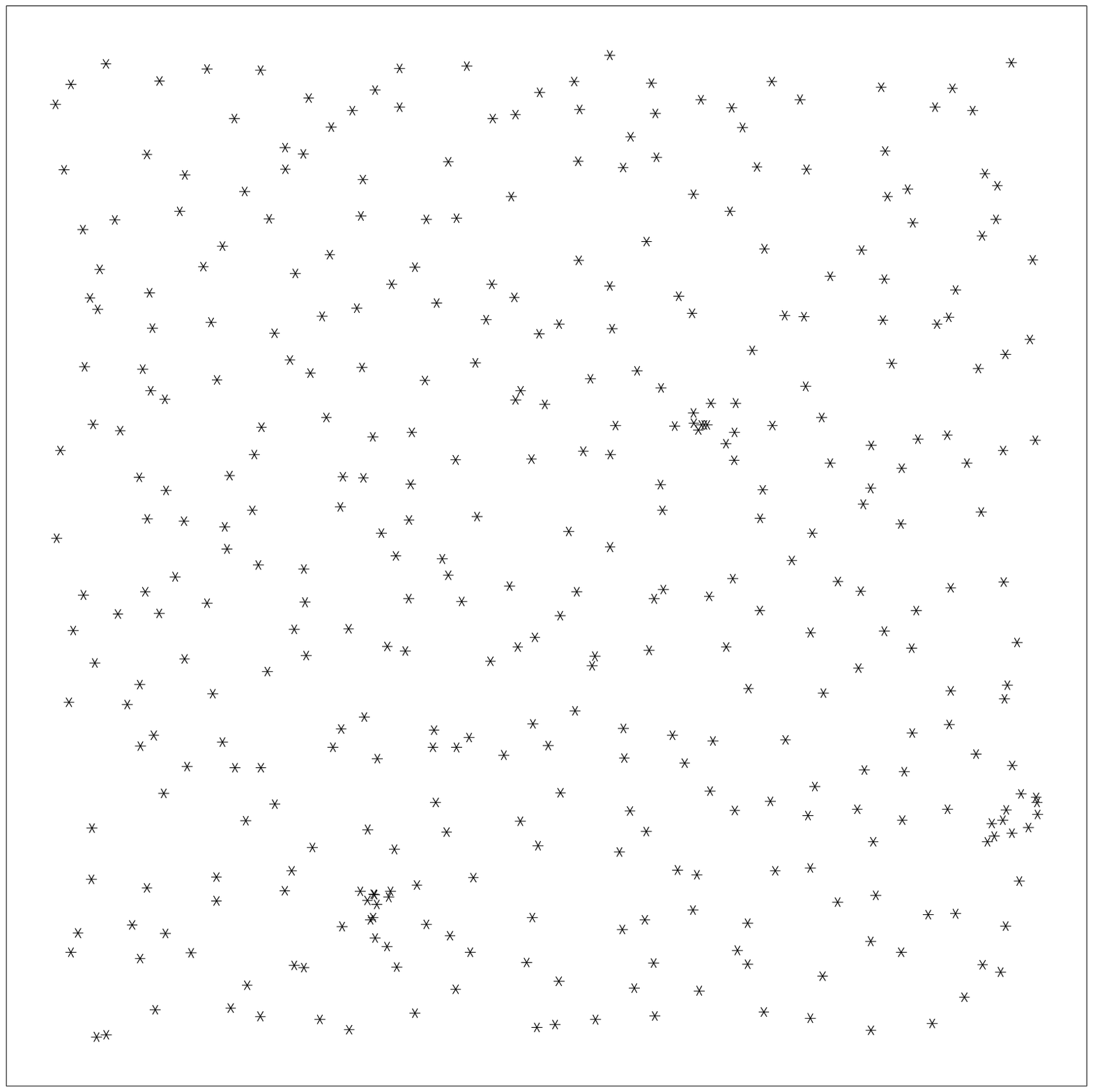} \end{minipage}
 \epsfxsize=3.9cm
 \begin{minipage}{\epsfxsize} \epsffile{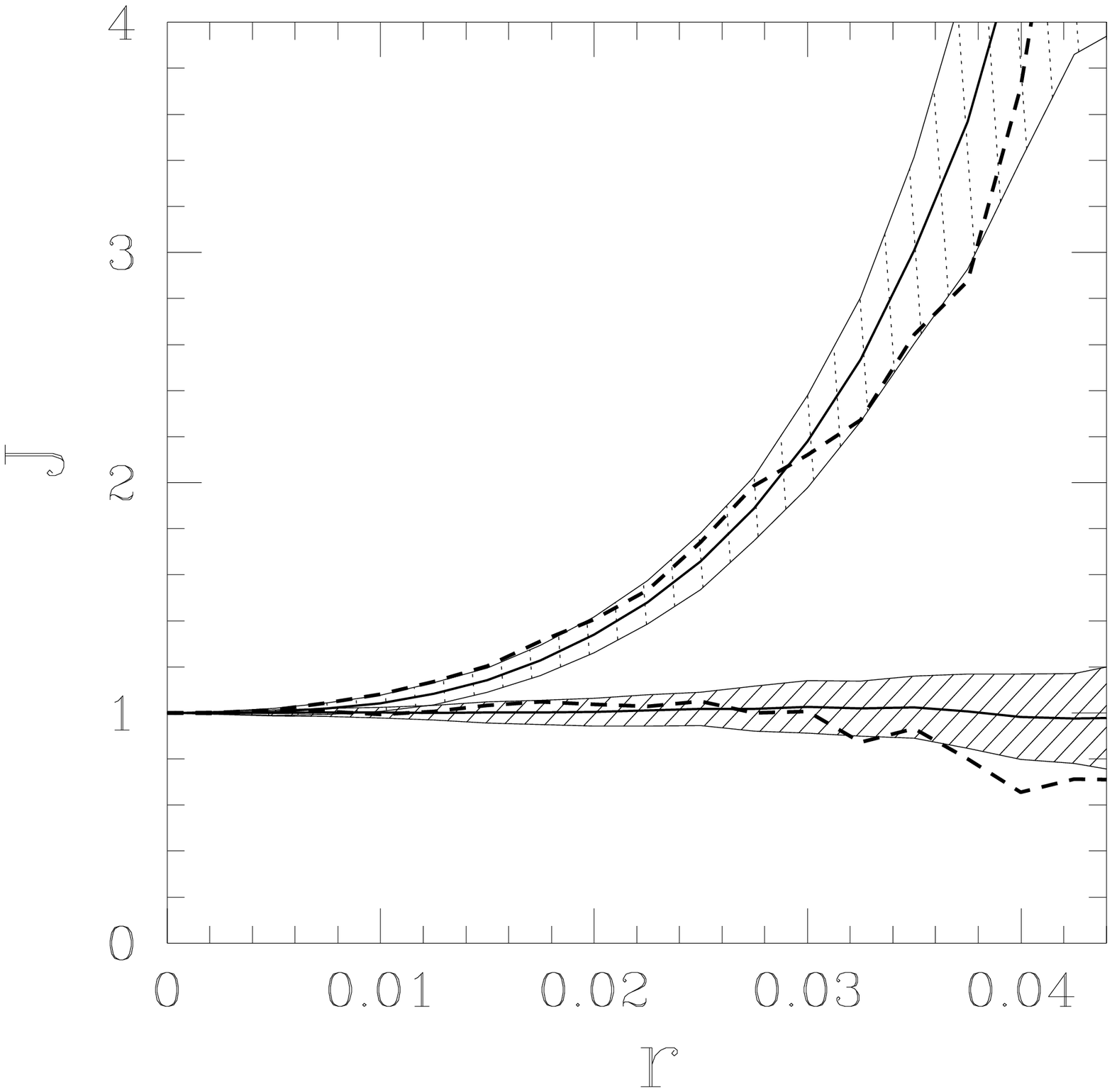} \end{minipage}
 \end{center}
\caption[]{\label{fig:2D}
On the left a realization of a Poisson process and in the middle a
realization of the regular example\protect\cite{baddeley:cautionary}
is displayed.
The plot on the right side shows the $J(r)$ for the Poisson process
(dark dashed) and for the regular example (light dotted). The areas
correspond to the 1$\sigma$ error estimated from fifty
realizations. The dashed lines are the $J(r)$ for the particular
realizations shown in the left and middle panels, $r$ is in units of
the side length of the square.}
\end{figure}
To characterize the clustering of a point process, van Lieshout and
Baddeley~{}\cite{vanlieshout:j} proposed to use the ratio
\begin{equation}
J(r) = \frac{1-G(r)}{1-F(r)} .
\end{equation}
Inherited from $F(r)$ and $G(r)$, $J(r)$ depends on correlations of
arbitrary order. For a homogeneous Poisson process we get $J(r) = 1$.

A process with enhanced clumping implies $J(r)\le1$, whereas regular
structures are indicated by $J(r)\ge1$.  Regular structures are seen
for instance in a periodic, or a crystal--like arrangement of points.
In a statistical sense, and opposed to clustering, regular
(``ordered'') structures are also seen in liquids. Qualitatively one
may explain the behavior of $J(r)$ in the following way:\\
$\bullet$ In a clustered distribution of points $G(r)$ increases more
rapidly with growing radius than for a Poisson process, since the
nearest neighbour is typically in the close surrounding. $F(r)$
increases more slowly, since an arbitrary point is typically in between
the clusters. These two effects give rise to a $J(r) \le 1$. \\
$\bullet$ In the contrary, in a regular process, $G(r)$ is lowered
with respect to a Poisson process since the nearest neighbour is
typically at a finite, in the case of a crystal, characteristic
distance. $F(r)$ is increasing stronger, since the typical distance
from a random point to a point on a regular structure is
smaller. These two effects cause a $J(r) \ge 1$.\\
$\bullet$ $J(r) = 1$ indicates the borderline between clustered and regular
structures.\\
We illustrate these properties with two different point processes: the
homogeneous Poisson process and the process constructed by Baddeley
and Silverman~{}\cite{baddeley:cautionary}.  Both processes have the
same two--point characteristics, i.e.\ $\xi=0$, but the example of
Baddeley and Silverman is regular by construction.
In Fig.~\ref{fig:2D} we display realizations of both processes with 
388 points in a square.  By visual inspection and with $J(r)\ge1$, the 
process given by Baddeley and Silverman shows regular structures, 
moreover $J(r)$ clearly distinguishes between these two random 
processes which both have $\xi(r)=0$. Obviously the differences in 
$J(r)$ result from high--order correlations only.

\section{Supercluster data}

\begin{figure}
 \begin{center}
 \epsfxsize=7.1cm
 \begin{minipage}{\epsfxsize} \epsffile{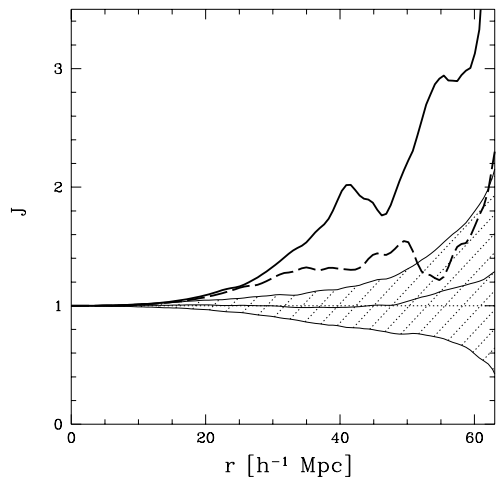} \end{minipage} 
 \end{center}
\caption[]{\label{fig:scJ} $J(r)$ for the supercluster catalogue
(southern part: solid line, northern part: dashed line), and the
1--$\sigma$ area of $J(r)$ for a Poisson Process with the same number
density determined from 100 realizations.}
\end{figure}

We investigate the supercluster
sample~{}\cite{einasto:supercluster_data} within galactic latitude
$|b|>20^\circ$ and a maximum radial distance of 330\hMpc, and perform
our analysis separately for the northern and southern parts.  
To estimate the $F(r)$ and $G(r)$ from the data we use the reduced
sample estimators~{}\cite{kerscher:regular} thereby taking care of edge
effects.
As seen in Fig.~\ref{fig:scJ}, $J(r)$ for the superclusters is clearly
above one, indicating regular structures.  With a nonparametric Monte
Carlo test we show~{}\cite{kerscher:regular} that these results are
incompatible with a random configuration of points at confidence level
of 95\%. However, the superclusters were identified with a
friend--of--friend procedure~{}\cite{einasto:supercluster_data} from
the Abell/ACO cluster sample.  Applying a friend--of--friend procedure
to a random set of points we recognize with the $J(r)$ that the
regularity seen in the distribution of superclusters in the northern
part (galactic coordinates) may be a spurious result of the
friend--of--friend algorithm. Still, the regularity seen with the
$J(r)\ge1$ in the southern part, cannot be explained with random
points and a friend--of--friend procedure afterwards.  Future
investigations will clarify the level of significance of this
regularity~{}\cite{kerscher:regular}.

The difference between the northern and southern parts are not only
due to the different selection criteria of the Abell and ACO parts,
since large fluctuations (i.e.\ cosmic variance) are also seen in the
galaxy distribution up to scales of
200\hMpc~{}\cite{kerscher:fluctuations}.

\section*{Acknowledgments}

I would like to thank Maret Einasto and coworkers fo making their
supercluster catalogue available to the public. Especially the
discussions with Herbert Wagner clarified on the subject and helped a
lot.

\section*{References}

\end{document}